# Essential finite-size effect in the 2D XY model

S.G. Chung

Department of Physics, Western Michigan University, Kalamazoo, MI 49008

Abstract

The thermodynamics of the 2D XY model is formulated by a transfer matrix method and analyzed by a density matrix renormalization group. The finite-size scaling and the beta function of the model are studied by the Roomany-Wyld renormalization group theory. It is found that the 2D XY model has an essential finite-size effect and the Berezinskii-Kosterlitz-Thouless transition with the critical temperature $T_{BKT} = 0.892$ appears in a finite system of 2000 - 3000 spins as a massless to massive transition with the effective critical temperature $T_c = 1.07 \pm 0.01$.

The 2D XY model describes classical planar spins with nearest neighbor interactions. The Hamiltonian is

$$H_{XY} = -\sum_{<ij>} \cos(\theta_i - \theta_j) \qquad (1)$$

This model has been intensively studied as a model which undergoes the Berezinskii-Kosterlitz-Thouless (BKT) transition,[1] the binding-unbinding transition of a vortex-antivortex pair that coexists with spin waves below the transition temperature. The remarkable point of the BKT theory is that it predicts an essential singularity,[2] an exponential growth of the correlation length and other thermodynamic quantities near the transition in contrast to the power-law behavior in a 2-nd order transition. This prediction has been tested numerically and reported confirmed again and again.[3] However, there are still some legitimate questions. In fact, there are 3 or 4 unknown (fitting) parameters in the theory. Moreover, it was pointed out that critical region to which the renormalization group (RG) equations confidently apply is very narrow, $(T-T_{BKT})/T_{BKT} < 10^{-2}$.[4] However, due to a strong finite-size effect, log L (L is system size) dependence of quantities, approaching such narrow temperature region by standard Monte Carlo (MC) simulations is almost impossible.[3]

Three methods have been proposed for handling this finite-size problem. One is Olsson's self-consistent boundary condition method.[5] This method can virtually eliminate finite-size fluctuations. Another is the matching method of Hasenbusch and Pinn.[6] This method is based on the exact solution available for the BCSOS (body-centered solid-on-solid) model. Comparing the block-spin RG flow of the XY model with that of the BCSOS model at long distance (the two models belong to the same universality class), the method introduces systematic errors which decay as $L^{-2}$, thereby overcoming the strong finite-size problem. These two methods remarkably agree on the critical temperature, giving $T_{BKT} = 0.892$. The third and recent attempt is a short-time dynamic approach due to Zheng, Schulz and Trimper.[7] Since the critical exponents at equilibrium enter into the short-time dynamic scaling, and the non-equilibrium spatial correlation length is small in the short-time dynamic evolution, the method can handle lower temperatures, up to 0.94 compared to 0.98 of canonical MC simulations. The critical temperature thus obtained 0.894 is very close to the seemingly exact value 0.892.

In the above, the thermodynamics limit is discussed. In practice, however, when certain physical systems are considered by the 2D XY model, the strong finite-size effect is essential. In fact, Bramwell and Holdsworth (BH)[8] studied this issue in the context of layered Heisenberg magnets with planar anisotropy which are well regarded as quasi-2D XY systems.[9] In these layered magnets, there is a very sharp crossover from a 3D behavior to a 2D fluctuation when lowering temperature. Moreover, the intra-layer exchange coupling J and inter-layer one J′ have a typical ratio $J/J' \sim 10^3$ to $10^4$, and therefore only fluctuation of length scale less than the order of $(J/J')^{1/2}$ = 30 ~ 100 are two dimensional. These layered magnets are thus well described as a finite-size, $10^3$ ~ $10^4$ spins or sites, 2D XY model. Using the linearized RG equations for finite-size scaling,[10] BH found an effective transition temperature $T_C$ = 1.080 ± 0.004 for 1024 spins and $T_c \sim 1.02$ for $10^4$ spins, and a power-law behavior of magnetization with critical exponent β = 0.23, explaining some experimental findings.[9] A remarkable point is that β = 0.23 is universal, irrespective of lattice types, spin values and degree of planar anisotropy of layered magnets, and claimed to be a signature of 2D XY behavior. While experiments are nicely explained, the BH theory may be questionable in the following points. First, it relies on the BKT theory that the correlation length behaves as,[2]

$$\xi = \exp\left(\frac{\pi}{\{c(T-T_{BKT})\}^{1/2}}\right) \qquad (2)$$

where c is a constant of order 2. However, strictly speaking,[4] it is correct only in a narrow temperature window, $(T - T_{BKT})/T_{BKT} < 10^{-2}$, and its use outside this region needs a justification. The effective temperature $T_C$ is defined in BH as the temperature when $\xi = L$. Second, the "shifted" temperature

$$T^*(L) \approx T_{BKT} + \frac{\pi^2}{4c(\ell nL)^2} \tag{3}$$

plays an important role in BH. This is the temperature where the renormalized spin-wave stiffness in the finite system takes the universal value $2/\pi$. This is determined by Monte Carlo as the temperature where the magnetization becomes a spin-wave estimate at $T = T_{BKT}$,

$$M = \left(\frac{1}{2L^2}\right)^{1/16} \tag{4}$$

The argument seems less precise and again uses the BKT theory beyond its confirmed validity.

In this paper, we study the finite-size effect in the 2D XY model by a transfer matrix (TM) method[11] with the help of the density matrix RG (DMRG)[12] technology and Roomany-Wyld RG theory.[13] Our approach does not involve any uncontrolled parameters, nor rely on the BKT theory. It demonstrates that the finite-size XY model with 2000 - 3000 spins behaves precisely like a system with the transition temperature $T_C = 1.07 \pm 0.01$ separating the massless and massive phases in consistent with the $T_c = 1.080 \pm 0.004$ in the BH theory for 1024 spins.

To calculate the partition function

$$Z = \int \prod_{i=1}^{N} d\theta_i \exp(-\beta H_{XY}) \tag{5}$$

where $\theta_i = (\theta_{i1}, \theta_{i2}, \ldots, \theta_{iM})$, consider a ribbon geometry; M sites in the width direction and N sites along the long, circular direction. With a negligible error at the edges, the Boltzmann factor in (5) can be written as

$$\prod_{i=1}^{N} K(\boldsymbol{\theta}_i, \boldsymbol{\theta}_{i+1}) \tag{6}$$

with the TM operator

$$K(\boldsymbol{\theta}_i, \boldsymbol{\theta}_{i+1}) = \prod_{j=1}^{M-1} \exp\left[\frac{\beta}{2}\left\{\cos(\theta_{ij} - \theta_{i+1j}) + \cos(\theta_{ij} - \theta_{ij+1}) + \cos(\theta_{i+1j} - \theta_{i+1j+1}) + \cos(\theta_{ij+1} - \theta_{i+1j+1})\right\}\right] \tag{7}$$

Now write

$$\int \prod_{i=1}^{N} d\boldsymbol{\theta}_i = \int \prod_{i=1}^{N+1} d\boldsymbol{\theta}_i \delta(\boldsymbol{\theta}_1 - \boldsymbol{\theta}_{N+1}) \tag{8}$$

and expand as

$$\delta(\boldsymbol{\theta}_1 - \boldsymbol{\theta}_{N+1}) = \sum_n \psi_n^*(\boldsymbol{\theta}_{N+1})\psi_n(\boldsymbol{\theta}_1) \tag{9}$$

where $\{\psi_n\}$ is an orthonormal complete set. We choose $\{\psi_n\}$ such that, the TM equation,

$$\int_0^{2\pi} d\boldsymbol{\theta} K(\boldsymbol{\theta}, \boldsymbol{\theta}')\psi_n(\boldsymbol{\theta}) = \exp(-\varepsilon_n)\psi_n(\boldsymbol{\theta}') \tag{10}$$

With these manipulations and using the TM equation repeatedly, we arrive at

$$Z = \sum_n \exp(-\varepsilon_n N) = \exp(-\varepsilon_0 N), \tag{11}$$

where $\varepsilon_0$ is the lowest energy. The last equality holds for N>>M>>1. The free energy per site is given by

$$f = T\frac{\varepsilon_0}{M} \tag{12}$$

The correlation function can be evaluated likewise,[11]

$$\langle \theta(0)\theta(r) \rangle = \langle \psi_0 | \theta(0)\theta(r) | \psi_0 \rangle \tag{13}$$

where $\psi_0$ denotes the eigenvector for the eigenvalue $\varepsilon_0$. The problem is thus reduced to the TM eigenvalue problem (10).

To analyze the TM equation (10), we use DMRG recently developed for 1d quantum systems.[12] In fact, the TM equation contains the Hamiltonian in a Boltzmann factor, so it is essentially a quantum M body problem. The TM-DMRG method is now a powerful tool for studying the thermodynamic properties of 1d quantum systems.[14] In the latter case, the Suzuki-Trotter decomposition of the Boltzmann factor leads to a non-Hermitian TM equation, requiring a little bit of numerical innovations.[11,15] In the present case, the TM operator K is real and symmetric, and less troublesome. On the other hand, unlike the quantum 1d cases where the TM equations are discrete from the outset, our TM equation is an integral equation, and needs a descretization. For that purpose, we use a Gaussian integration formula, transforming (10) to

$$\pi^M \sum_{ijk...=1}^{p} w_i w_j w_k ... K(ijk... : i'j'k'...) \psi_n(ijk...) = \exp(-\varepsilon_n) \psi_n(i'j'k'...) \qquad (14)$$

where $\psi_n$ (ijk…) means $\psi_n(\theta_i \theta_j \theta_k ...)$ and $\theta_i = \pi(x_i+1)$ etc with $w_i$ and $x_i$ being the Gaussian abscissas and weight factors.[16] If we rescale as

$$(\pi^M w_i w_k w_k ...)^{1/2} \psi(ijk...) \to \psi(ijk...) \qquad (15)$$

then the eigenvector is Euclid normalized and the TM equation becomes

$$\sum_{ijk...} OS(ii')W(ij:i'j')OS(jj')W(jk:j'k')OS(kk')...\psi(ijk...) = \exp(-\varepsilon_n)\psi(i'j'k'...) \qquad (16)$$

where W(ij:i′j′) denotes the Boltzmann factor on the right hand side of (7) with i+1=i′ and j+1 = j′, and OS(ii′) = $\pi(w_i w_i')^{1/2}$ etc.

The DMRG procedure to get the lowest eigenvalues $\varepsilon_n$ is as follows. Step 1, consider 3 sites, that is the $p^3 \times p^3$ eigenvalue problem. Note that, unlike the 1d quantum cases where the degree of freedom at each site is small, the degree of freedom here, p, needs to be large, so we have to use the 3 sites algorithm whose accuracy was discussed before.[17] Step 2, construct the density matrix out of the ground state $\psi_0$,

$$\rho(i'j':ij) = \sum_k \psi_0^*(i'j'k)\psi_0(ijk), \qquad (17)$$

diagonalize it and retain the largest m states. Denote the m eigenvectors by $\mathbf{X_1}, \mathbf{X_2}, ..., \mathbf{X_m}$ and eigenvalues by $\xi_1, \xi_2, ..., \xi_m$. So $(\mathbf{X_1 X_2 ... X_m}) \equiv O(ij:\xi)$ constitutes a $p^2 \times m$ transformation matrix for the change of basis $\theta \times \theta \to \xi$. The renormalized TM operator is

$$\sum_{ii'jj'} O^*(\xi':i'j')OS(ii')W(ij:i'j')OS(jj')W(jk:j'k')O(ij:\xi) = T_2(\xi'k':\xi k) \qquad (18)$$

Symbolically • W • W → $T_2$ where the solid circles represent OS. Step 3, go back to step 1 with the renormalized operator $T_2$ • $T_2$ for the 5 sites, and diagonalize the resulting $pm^2 \times pm^2$ eigenvalue problem to get the new ground state $\psi_0$. Step 4, repeat the step 2 to construct the reduced density matrix, diagonalize it to get a new transformation matrix O and then a RG transformation, $T_2$ • W → $T_3$. The 7 sites system is now represented by $T_3$ • $T_3$. Repeating the procedure, one can get a larger and larger system.

One needs to keep track of, in the course of RG transformations, the spin operator (cos θ, sin θ) at each site to calculate various expectation values such as the correlation function (13). For example in the RG operation •W•W→$T_2$ in the steps 1 and 2, the angle $\hat{\theta}$ (caret means an operator) in the middle site is a one-particle operator, diagonal in the original angle representation, and therefore represented as

$$(\hat{1} \times \hat{\theta})_{\theta_i \theta_j : \theta_{i'} \theta_{j'}} = \delta_{ii'} \theta_j \delta_{jj'} \qquad (19)$$

Thus the RG transformation $\hat{\theta} \to \hat{\theta}_{new}$ is

$$\sum_{ii'jj'} O(\xi':i'j')\delta_{ii'} \theta_j \delta_{jj'} O(ij:\xi) \equiv \hat{\theta}_{new}(\xi'\xi) \qquad (20)$$

Let us first check the accuracy of our TM-DMRG method. For this purpose, we calculate the internal energy per site

$$E = -T^2 \frac{\partial}{\partial T}(\varepsilon_0 / M) \tag{21}$$

It is now well-established that the ground state energy per site $\varepsilon_0/M$ in the thermodynamic limit can be easily obtained by measuring the energy increase in successive steps of the DMRG procedure. We have started the calculation with p = m = 12 and checked the complete convergence up to p = m = 24. In the T = 0.1 case, we have checked the convergence by calculating up to the p = m = 32 case. The result is given in Fig. 1 where the kinetic energy T/2 is added to compare with the latest simulation on the kinetic XY model and the MC calculations.[18] While agreement with MC results is good, the TM-DMRG result is about 10% larger than the kinetic simulation result at T > 1.4. Since the finite-size effect is negligibly small at higher T, the discrepancy is a puzzle at the moment. However, it is noted that similar discrepancies have been observed for the specific heat among MC calculations[3].

We now calculate not only the smallest $\varepsilon_0$ but also the next smallest $\varepsilon_1$. In fact one can show easily that the correlation function (13) behaves for r >>1 like

$$\sim \exp\{-(\varepsilon_1 - \varepsilon_0)r\} \tag{22}$$

Thus Gap (L) ≡ $\varepsilon_1$- $\varepsilon_0$ measures the gap energy of the system at size L. Then due to the Roomany-Wyld RG theory,[13] the effective critical temperature $T_c$ of the finite-size 2D XY model can be located by calculating L x Gap (L) vs T. Fig. 2 shows the result for L = 37, 43, 49, 55 and 61 (1500 ~ 3500 spins) using p = 12 and m = 60. The cases p = 16, m = 60 and p = 12, m = 66 give less than 1% correction to the result. We have also calculated the Roomany-Wyld approximate for the beta function

$$\beta_{LL'}^{RW}(T) = \frac{1 + \ell n(G_L / G_{L'}) / \ell n(L / L')}{(D_L D_{L'} / G_L G_{L'})^{1/2}} \tag{23}$$

where $G_L = Gap(L)$ and $D_L = \partial G_L/\partial T$. Fig. 3 shows the result for L = 45, 47, 49 and 51 (2000 ~ 3000 spins) with L′ = L + 2. Again p = 12 and m = 60. Small drift of the zero point of the beta function towards higher $T_c$ for larger L may be due to an accumulation of the truncation (p x m → m) error in DMRG. Based on the results Fig. 2 and 3, the effective critical temperature of the finite XY system with 2000 - 3000 spins is determined to be $T_c = 1.07 \pm 0.01$, in consistent with the $T_c = 1.080 \pm 0.004$ in the BH theory[8] for 1024 spins.

To conclude, the thermodynamics of the 2D XY model is formulated by a transfer matrix method and analyzed by a density matrix renormalization group. The finite-size scaling and the beta function of the model are studied by the Roomany-Wyld renormalization group theory. It is found that the 2D XY model has an essential finite-size effect and the Berezinskii-Kosterlitz-Thouless transition with the critical temperature $T_{BKT} = 0.892$ manifests itself in the finite system of 2000 - 3000 spins as a massless to massive transition with the effective critical temperature $T_c = 1.07 \pm 0.01$.

This work was partially supported by NSF under DMR 980009N and utilized the SGI/CRAY Origin 2000 at the National Center for Supercomputing Applications at University of Illinois at Urbana-Champaign.

**Figure Captions**

Fig. 1  Internal energy vs temperature. Open circles are from the equation of motion method for the kinetic XY model and diamonds are MC data available up to T = 1.5.[18]  Solid circles are from the TM-DMRG.

Fig. 2  L x Gap (L) vs temperature for L = 37,43,49,55 and 61 from bottom to top. The result is for p = 12 and m = 60. The cases p = 12, m = 66 and p = 16, m = 60 give less than 1% correction to the result.

Fig. 3  Beta function vs temperature for L = 45,47,49 and 51 with L′ = L + 2 from top to bottom. Again p = 12 and m = 60. Together with Fig. 2, the effective critical temperature of the finite XY system with 2000 - 3000 spins is determined to be $T_C$ = 1.07 ± 0.01.

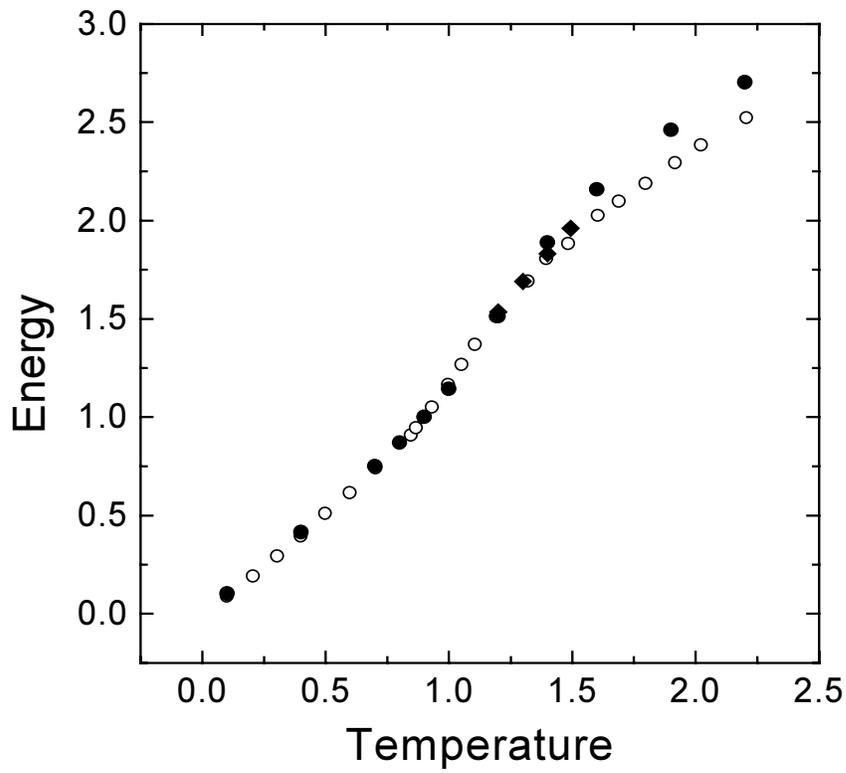

Fig 1 Chung

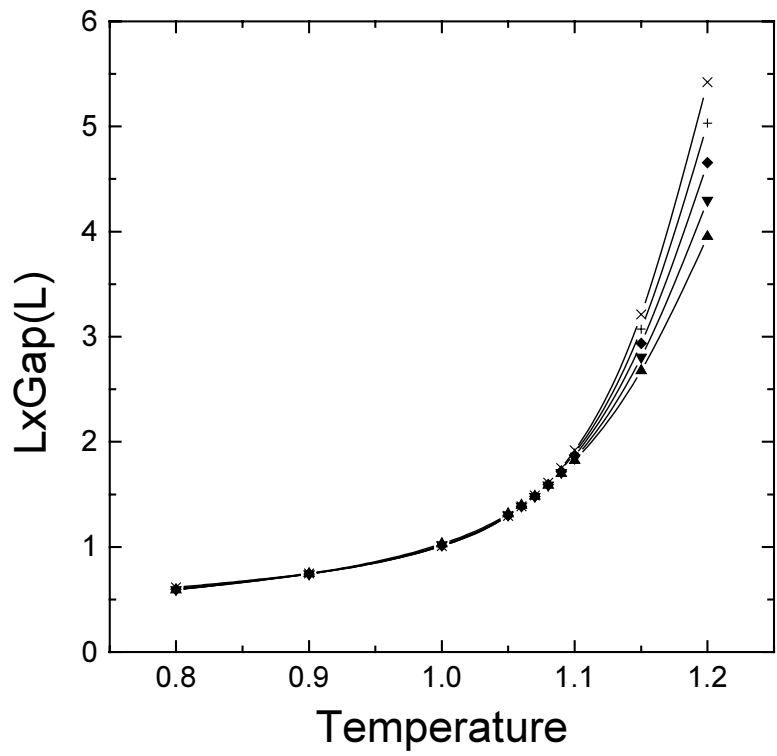

Fig 2 Chung

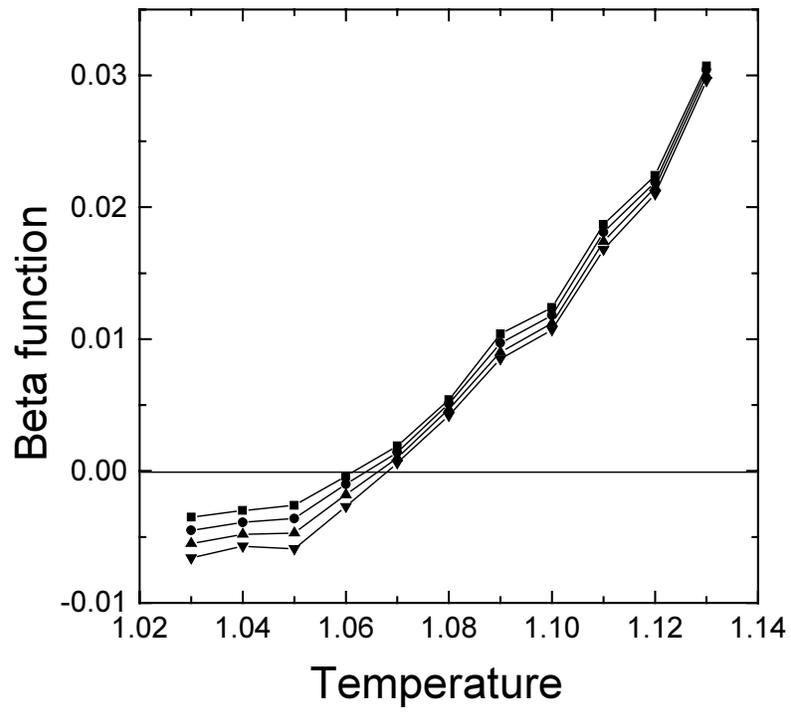

Fig 3 Chung